\documentclass[10pt,twocolumn]{article}

\usepackage{amsmath}
\usepackage{amssymb}
\usepackage{amsthm}
\usepackage{aligned-overset}
\usepackage{psfrag}
\usepackage{color}
\usepackage[usenames,dvipsnames,table]{xcolor}
\usepackage{units}
\usepackage[top=54pt, left=54pt, right=54pt, bottom=54pt, paper=letterpaper]{geometry}
\usepackage{paralist}
\usepackage[table]{xcolor}
\usepackage[noadjust]{cite}
\usepackage{booktabs}
\usepackage[format=plain,labelfont=it]{caption}
\usepackage[colorlinks=false]{hyperref}
\usepackage{url}
\usepackage{stmaryrd}
\usepackage{algpseudocode}
\usepackage{algorithm}
\usepackage{tabularx}
\usepackage{titlesec}
\usepackage{accents}
\newlength{\dhatheight}

\titleformat{\section}{\centering\normalfont\scshape}{\Roman{section}.}{5pt}{}
\titleformat{\subsection}{\normalfont\it}{\Alph{subsection}.}{5pt}{}
\titleformat{\subsubsection}{\normalfont\it}{\hspace{4mm}\arabic{subsubsection})}{5pt}{}

\newcommand\infoFootnote[1]{%
  \begingroup
  \renewcommand\thefootnote{}\footnote{#1}%
  \addtocounter{footnote}{-1}%
  \endgroup}

\newtheorem{thm}{Theorem}
\newtheorem{lem}[thm]{Lemma}
\newtheorem{assum}{Assumption}
\newtheorem{spec}{Specification}

\newcommand{\R}{\mathbb{R}}

\newcommand{\N}{\mathbb{N}}

\newcommand{\eb}{\boldsymbol{e}}
\newcommand{\fb}{\boldsymbol{f}}

\newcommand{\rb}{\boldsymbol{r}}
\newcommand{\ub}{\boldsymbol{u}}

\newcommand{\xb}{\boldsymbol{x}}
\newcommand{\yb}{\boldsymbol{y}}
\newcommand{\zb}{\boldsymbol{z}}

\newcommand{\Sigmab}{\boldsymbol{\Sigma}}
\newcommand{\lambdab}{\boldsymbol{\lambda}}
\newcommand{\zerob}{\boldsymbol{0}}

\newcommand{\Ab}{\boldsymbol{A}}
\newcommand{\Bb}{\boldsymbol{B}}

\newcommand{\Hb}{\boldsymbol{H}}
\newcommand{\Ib}{\boldsymbol{I}}

\newcommand{\Qb}{\boldsymbol{Q}}
\newcommand{\Rb}{\boldsymbol{R}}

\newcommand{\Gb}{\boldsymbol{G}}

\newcommand{\Pb}{\boldsymbol{P}}
\newcommand{\varphib}{\boldsymbol{\varphi}}
\newcommand{\varepsb}{\boldsymbol{\varepsilon}}

\newcommand{\Hbt}{\tilde{\boldsymbol{H}}}
\newcommand{\Gbt}{\tilde{\boldsymbol{G}}}
\newcommand{\Dbt}{\tilde{\boldsymbol{D}}}
\newcommand{\fbt}{\tilde{\boldsymbol{f}}}
\newcommand{\ebt}{\tilde{\boldsymbol{e}}}
\newcommand{\Rbt}{\tilde{\boldsymbol{R}}}
\newcommand{\rbt}{\tilde{\boldsymbol{r}}}
\newcommand{\Ubt}{\tilde{\boldsymbol{U}}}
\newcommand{\Ybt}{\tilde{\boldsymbol{Y}}}
\newcommand{\Vbt}{\tilde{\boldsymbol{V}}}
\newcommand{\Sigmabt}{\tilde{\boldsymbol{\Sigma}}}

\newcommand{\Hbh}{\hat{\boldsymbol{H}}}
\newcommand{\Gbh}{\hat{\boldsymbol{G}}}
\newcommand{\fbh}{\hat{\boldsymbol{f}}}
\newcommand{\ebh}{\hat{\boldsymbol{e}}}
\newcommand{\Rbh}{\hat{\boldsymbol{R}}}
\newcommand{\rbh}{\hat{\boldsymbol{r}}}
\newcommand{\zbh}{\hat{\boldsymbol{z}}}

\newcommand{\Hbv}{\check{\boldsymbol{H}}}
\newcommand{\Gbv}{\check{\boldsymbol{G}}}
\newcommand{\fbv}{\check{\boldsymbol{f}}}
\newcommand{\ebv}{\check{\boldsymbol{e}}}
\newcommand{\Rbv}{\check{\boldsymbol{R}}}
\newcommand{\rbv}{\check{\boldsymbol{r}}}

\newcommand{\Kc}{\mathcal{K}}

\newcommand{\blind}[1]{\textcolor{white}{#1}}

\renewcommand{\boldsymbol}[1]{#1}
\renewcommand{\mathbf}[1]{\mathrm{#1}}

\DeclareMathOperator{\diag}{diag}

\def\tvdots{\vbox{\baselineskip=2pt \lineskiplimit=0pt \kern6pt \hbox{.}\hbox{.}\hbox{.}}}

\title{\vspace{-2mm}\bf
On the security of randomly transformed quadratic programs \\
for privacy-preserving cloud-based control${}^\ast$}
\author{Philipp Binfet, Nils Schl\"uter, and Moritz Schulze Darup\vspace{2mm}}
\date{}

\author{Philipp Binfet, Nils Schl\"uter, and Moritz Schulze Darup}

\begin{document}
\maketitle

\setlength{\abovedisplayskip}{8pt plus 2pt minus 1pt}  
\setlength{\belowdisplayskip}{8pt plus 2pt minus 1pt}  

\textbf{\textit{Abstract}.} {\bf%
Control related data, such as system states and inputs or controller specifications, is often sensitive. Meanwhile, the increasing connectivity of cloud-based or networked control results in vast amounts of such data, which poses a privacy threat, especially when evaluation on external platforms is considered.
In this context, a cipher based on a random affine transformation gained attention, which is supposed to enable privacy-preserving evaluations of quadratic programs (QPs) with little computational overhead compared to other methods.

This paper deals with the security of such randomly transformed QPs in the context of model predictive control (MPC).
In particular, we show how to construct attacks against this cipher and thereby underpin concerns regarding its security in a practical setting. To this end, we exploit invariants under the transformations and common specifications of MPC-related QPs.
Our numerical examples then illustrate that these two ingredients suffice to extract information from ciphertexts.
\infoFootnote{P.~Binfet, N.~Schl\"uter, and M.~Schulze Darup are with the \href{https://rcs.mb.tu-dortmund.de/}{Control and~Cyber-physical Systems Group}, Faculty of Mechanical Engineering, TU Dortmund University, Germany. E-mails:  \href{mailto:moritz.schulzedarup@tu-dortmund.de}{\{philipp.binfet, nils.schlueter, moritz.schulzedarup\}@tu-dortmund.de}. \vspace{0.5mm}}
\infoFootnote{\hspace{-1.5mm}$^\ast$This paper is a \textbf{preprint} of a contribution to the 62nd IEEE Conference on Decision and Control 2023. © 2023 IEEE. Personal use of this material is permitted.  Permission from IEEE must be obtained for all other uses, in any current or future media, including reprinting/republishing this material for advertising or promotional purposes, creating new collective works, for resale or redistribution to servers or lists, or reuse of any copyrighted component of this work in other works.}%
}

\noindent

\section{Introduction}

The privacy-preserving evaluation of a control related functionality is the main focus of encrypted control. In this context, quadratic programs (QPs) are of special interest, because they serve as the corner stone for solutions in many decision-making problems, where model predictive control (MPC) is a famous example.
Roughly speaking, data privacy can be achieved via several cryptographic methods. Noteworthy are homomorphic encryption~\cite{chillotti2020tfhe,Cheon2017_CKKS}, secure multi-party computation~\cite{cramer2015secure}, and differential privacy~\cite{dwork2014algorithmic}. However, these methods come with specific drawbacks such as large overhead in terms of computation or communication or a privacy-accuracy trade-off. Nonetheless, QPs have been addressed with these methods, for example in~\cite{alexandru2020cloud,nozari2016differentially}.

A cipher which lately received attention is called random affine transformation (RT), random matrix encryption or affine masking. This method does not suffer from the aforementioned drawbacks
and it can be used for a confidential evaluation of optimization problems, e.g., on a cloud.
The application of RT ciphers to linear programming is addressed in~\cite{vaidya2009privacy,dreier2011practical,wang2011secure}, whereas QPs are considered in~\cite{Zhou2015QPoutsourcing,XuZhu2015}. A more general formulation for QPs can be found in~\cite{weeraddana2013per} and~\cite{sultangazin2020symmetries}, where the latter has a focus on linear MPC, as it is the case for~\cite{naseri2022mpctransform}. Finally, also nonlinear MPC~\cite{zhang2021privacy} and federated learning~\cite{hayati2022privacy} have been considered in this context.
Despite the fact that we collect these results under the umbrella of RT ciphers here, we point out that there are differences between them in terms of key reusage and additional permutations.

At this point, one may already suspect that the advantages of RTs do not come for free.
In order to confirm that intuition, this paper deals with the construction of concrete attacks on the RTs applied to MPC, where QPs are solved sequentially. For the confidential evaluation of these QPs, we consider RTs over real numbers (as in the literature above) and allow for different keys in every problem instance. An extension with additional permutations is shown later. This way, many of the publications which use RTs are addressed.
Now, as pointed out in~\cite{IFAC2023_RandomTrafo}, the ``vector-ciphertexts'' resulting from such an RT can be made secure. However, applying this cipher to QPs results in transformed QPs with slightly different ``matrix-ciphertexts'' for the parameters (such as the Hessian), which contain invariants.
We show that certain information is inevitably leaked from ciphertexts. Furthermore, based on the invariants in combination with ciphertexts and little additional knowledge (justified by Kerckhoffs' principle) one can break the cipher entirely. Our findings are illustrated by an application to setpoint and tracking MPC problems.

In the remainder of the paper, we first specify the transformed/encrypted QP and the corresponding RT (Section~\ref{sec:problemstatement}). Then, Section~\ref{sec:problemAnalysis} analyzes the peculiarities of these QPs on which the attacks in Section~\ref{sec:Attacks} are based.

\section{Problem statement}
\label{sec:problemstatement}

We are interested in solving various instances of the QP
\begin{equation}
\label{eq:originalQP}
  \zb_k^\ast :=\arg \min_{\zb}  \frac{1}{2} \zb^\top \Hb_k \zb + \fb_k^\top \zb \quad \text{s.t.} \quad \Gb_k \zb \leq \eb_k
\end{equation}
where $k\in\N$ distinguishes the (time-) varying parameters
\begin{equation}
\label{eq:parametersQP}
\Hb_k \in \R^{l\times l}, \quad\! \Gb_k \in \R^{q\times l}, \quad\! \fb_k \in \R^{l}, \quad\! \text{and} \quad\! \eb_k \in \R^{q}
\end{equation}
with constant dimensions $l,q\in \N$. Further,
we want to outsource this optimization to a cloud.
Here, the (honest-but-curious) cloud represents any external computation platform that is interested in learning the QP parameters $\Hb_k$, $\Gb_k$, $\fb_k$, and $\eb_k$ as well as the optimizer $\zb_k^\ast$ but executes computations as specified.
In order to establish privacy, it has been proposed in the literature (e.g., \cite{sultangazin2020symmetries}) to use an RT of the optimization variable according to
\begin{equation}
\label{eq:relationOfOptimizers}
    \zb = \Rb_k \yb +\rb_k,
\end{equation}
where an invertible matrix $\Rb_k \in \R^{l \times l}$ and a vector $\rb_k \in \R^{l}$ are randomly chosen for each $k$. With this at hand, one can transmit the transformed parameters
\begin{subequations}
 \label{eq:transformations}
\begin{align}
\label{eq:transformationsHG}
    \Hbt_k&:=\Rb_k^\top \Hb_k \Rb_k, & \Gbt_k&:=\Gb_k \Rb_k,\\
    \label{eq:transformations_fe}
    \fbt_k&:= \Rb_k^\top (\fb_k +\Hb_k \rb_k),& \ebt_k&:=\eb_k-\Gb_k \rb_k.
\end{align}
\end{subequations}
to the cloud, which then solves
\begin{equation}
\label{eq:transformedQP}
  \yb_k^\ast :=\arg \min_{\yb}  \frac{1}{2} \yb^\top \Hbt_k \yb + \fbt_k^\top \yb \quad \text{s.t.} \quad \Gbt_k \yb \leq \ebt_k
\end{equation}
instead of \eqref{eq:originalQP}.
Finally, based on the returned optimizer $\yb_k^\ast$, we easily recover the desired optimizer $\zb_k^\ast$ via~\eqref{eq:relationOfOptimizers}.
Correctness of this scheme can be easily verified (cf.~\cite[Thm.~1]{XuZhu2015}).
However, the question of interest in this paper is whether the privacy of the original QP parameters~\eqref{eq:parametersQP} (and optimizers~$\zb_k^\ast$) is indeed protected.

In this context, many authors conclude that~\eqref{eq:transformations} establishes privacy of~\eqref{eq:parametersQP} from the following observation. Assume the cloud came up with a guess $\Hbh_k$, $\Gbh_k$, $\fbh_k$, $\ebh_k$, $\Rbh_k$, and $\rbh_k$ for the original QP parameters
that is consistent with the transformations~\eqref{eq:transformations} and the data available to the cloud in terms of $\Hbt_k$, $\Gbt_k$, $\fbt_k$, and $\ebt_k$, i.e.,
\begin{subequations}
\label{eq:consistentGuess}
\begin{align}
\label{eq:consistentGuessHG}
    \Hbt_k&=\Rbh_k^\top \Hbh_k \Rbh_k, & \Gbt_k&=\Gbh_k \Rbh_k,\\
    \label{eq:consistentGuess_fe}
    \fbt_k&= \Rbh_k^\top (\fbh_k +\Hbh_k \rbh_k),& \ebt_k&=\ebh_k-\Gbh_k \rbh_k.\\[-6mm]
    \nonumber
\end{align}
\end{subequations}
Then, the cloud could easily generate infinitely many additional consistent guesses $\Hbv_k$, $\Gbv_k$, $\fbv_k$, $\ebv_k$, $\Rbv_k$, and $\rbv_k$ by choosing any invertible matrix $\Rbt_k \in \R^{l \times l}$ and any vector $\rbt_k \in \R^{l}$ and by specifying the additional guesses via
\begin{subequations}
\label{eq:additionalGuesses}
\begin{align}
\label{eq:additionalHG}
    \Hbv_k&:=\Rbt_k^\top \Hbh_k \Rbt_k, & \Gbv_k&:=\Gbh_k \Rbt_k,\\
    \label{eq:additional_fe}
    \fbv_k&:= \Rbt_k^\top (\fbh_k +\Hbh_k \rbt_k),& \ebv_k&:=\ebh_k-\Gbh_k \rbt_k,\\
    \label{eq:additionalRr}
    \Rbv_k&:=\Rbt_k^{-1} \Rbh_k, &  \rbv_k&:=\Rbt_k^{-1} ( \rbh_k-\rbt_k).\\[-6mm]
    \nonumber
\end{align}
\end{subequations}
Clearly, \eqref{eq:additionalHG} and \eqref{eq:additional_fe} just reflect another transformation of the form~\eqref{eq:transformations} using the chosen $\Rbt_k$ and $\rbt_k$. The relations~\eqref{eq:additionalRr} provide suitable updates of the guessed transformations.
Now, \eqref{eq:consistentGuess} and \eqref{eq:additionalGuesses} imply that even if one can solve the nonlinear system of equations underlying the transformations~\eqref{eq:transformations}, there exist infinitely many other solutions, and it seems impossible to select the one that actually applies. In other words, privacy stems from ambiguity of the applied transformation.
We will double-check this argumentation for popular problem specifications arising in model predictive control in the following.

\subsection{Problem specifications for predictive control}

Classical MPC leads to a sequence of optimal control problems in the form~\eqref{eq:originalQP}, where $\Hb_k$ and $\Gb_k$ are constant (see, e.g., \cite[Eq.~(7)]{Bemporad2002}).
This
is summarized in the following specification for later reference.
\begin{spec}
\label{spec:HGinvariant}
The matrices $\Hb_k$ and $\Gb_k$ in~\eqref{eq:originalQP} are constant,
i.e.,  $\Hb_k=\Hb_0$ and $\Gb_k=\Gb_0$ for every $k\in\N$.
\end{spec}
Furthermore, the optimization variable $\zb$ typically represents the predicted input sequence in MPC, which is, among other constraints, often subjected to box constraints of the form
\begin{equation}
\label{eq:zBoxConstraints}
\underline{\zb} \leq \zb \leq \overline{\zb}.
\end{equation}
Such or similar constraints lead to constant parts in $\eb_k$, as specified next.
\begin{spec}
\label{spec:ekFixedUpperPart}
At least the first $q_{\text{fix}} \in \N$ elements of the constraint vectors $\eb_k$ are constant,
i.e.,
\begin{equation}
\label{eq:ekFixedUpperPart}
\eb_k = \begin{pmatrix}
    \eb^{\text{fix}} \\
    \eb_k^{\mathrm{var}}
\end{pmatrix} \quad \text{for every}\,\, k \in \N
\end{equation}
with $\eb^{\text{fix}} \in \R^{q_{\text{fix}}}$ and $\eb_k^{\mathrm{var}} \in \R^{q_{\text{var}}}$, where $q_{\text{var}}:=q-q_{\text{fix}}$.
\end{spec}

Finally, for classical MPC, the vectors $e_k$ and $f_k$ are affine in the current system state $x_k$ \cite[Eq.~(7)]{Bemporad2002}.
Hence, $e_k$ and $f_k$ are (almost) constant whenever the states $x_k$ are (almost) constant. The latter applies, for instance, if the state converges to a setpoint. Thus, we also consider the following specification, where we assume exactly constant vectors for analysis purposes.
\begin{spec}
\label{spec:efkConstant}
The vectors $\fb_k$ and $\eb_k$ in~\eqref{eq:originalQP} are constant for every $k$ in some set $\Kc \subseteq \N$, i.e., $\fb_j = \fb_{k}$ and $\eb_j = \eb_{k}$ for all $j,k\in \Kc$.
\end{spec}

We will investigate the implications of Specifications~\ref{spec:HGinvariant}, \ref{spec:ekFixedUpperPart}, and \ref{spec:efkConstant} on the desired privacy in the upcoming sections.
Before doing so, we present a variant of the transformations in~\eqref{eq:transformations} which can increase the security.

\subsection{Additional random permutations}
\label{subsec:Permutations}

Inspired by \cite[Sect.~3.2]{salinas2016efficient}, we will also briefly discuss the combination of the transformations~\eqref{eq:transformations} with random permutations of the inequality constraints in~\eqref{eq:originalQP}.
More precisely, for every $k\in \N$,
a permutation matrix $\Pb_k \in \{0,1\}^{q \times q}$
(in addition to $\Rb_k$ and $\rb_k$) is randomly chosen which results in
\begin{equation}
\label{eq:permutedTransformations}
\Gbt_k^\prime :=\Pb_k \Gb_k \Rb_k \quad \text{and} \quad  \ebt_k^\prime :=\Pb_k (\eb_k - \Gb_k \rb_k)
\end{equation}
instead of the corresponding transformations in~\eqref{eq:transformations}. Accordingly, we substitute the constraints $\Gbt_k \yb \leq \ebt_k$ in~\eqref{eq:transformedQP} with $\Gbt_k^\prime \yb \leq \ebt_k^\prime$.
Note that, in light of such permutations, the assumed order of the constant and varying
parts of $\eb_k$ in~\eqref{eq:ekFixedUpperPart} is without loss of generality.

\section{Problem analysis}
\label{sec:problemAnalysis}

Before discussing possible attacks against the presented scheme, we analyze the transformed QP and the data available to the cloud. To avoid technicalities, we make the following assumption throughout the remaining paper.
\begin{assum}
\label{assum:HposDefGRankl}
The matrices $\Hb_k$ are positive definite and the rank of $\Gb_k$ is $l$ for every $k \in \N$.
\end{assum}

\subsection{Invariants and their relation to the dual problem(s)}

We begin by pointing out two invariants regarding the transformed parameters~\eqref{eq:transformations}.
These are closely related to the dual of the QP~\eqref{eq:originalQP}, which is well-known to be
\begin{equation}
\label{eq:dualQP}
    \!\!\lambdab_k^\ast=\arg\min_{0\leq \lambdab} \frac{1}{2}\lambdab^\top\! \Gb_k \Hb_k^{-1} \Gb_k \lambdab + ( \Gb_k \Hb_k^{-1} \fb_k\! + \eb_k)^\top\!\lambdab.\!
\end{equation}
Interestingly, the duals of~\eqref{eq:originalQP} and~\eqref{eq:transformedQP} are equal.
In fact, this immediately follows from the invariants summarized next.
\begin{lem}
The transformed parameters~\eqref{eq:transformations} are related to the original parameters via
    \begin{subequations}
    \label{eq:invariants}
\begin{align}
\label{eq:relationGHG}
    \tilde{\Gb}_k \tilde{\Hb}_k^{-1} \tilde{\Gb}_k^\top  &= \Gb_k \Hb_k^{-1} \Gb_k^\top \qquad \text{and} \\
     \label{eq:relationGHfe}
    \tilde{\Gb}_k \tilde{\Hb}_k^{-1} \tilde{\fb}_k + \tilde{\eb}_k
   &= \Gb_k \Hb_k^{-1} \fb_k + \eb_k\\[-6.5mm]
   \nonumber
\end{align}
    \end{subequations}
for any choice of $\Rb_k$ and $\rb_k$.
\end{lem}
\begin{proof}
   The invariants immediately follow from substituting the expressions~\eqref{eq:transformations} for $\tilde{\Hb}_k$, $\tilde{\Gb}_k$, $\tilde{\fb}_k$, and $\tilde{\eb}_k$.
\end{proof}
Moreover, the invariants~\eqref{eq:invariants} are closely related to Specifications~\ref{spec:HGinvariant} and~\ref{spec:efkConstant} as detailed next.
\begin{lem}
\label{lem:SpecsAndInvariants}
    If Specification~\ref{spec:HGinvariant} applies, then $\tilde{\Gb}_k \tilde{\Hb}_k^{-1} \tilde{\Gb}_k^\top$ is constant for every $k\in \N$. Further, if Specifications~\ref{spec:HGinvariant} and~\ref{spec:efkConstant} apply, then $\tilde{\Gb}_k \tilde{\Hb}_k^{-1} \tilde{\fb}_k + \tilde{\eb}_k$ is constant for every $k\in \Kc$.
\end{lem}

\begin{proof}
 Specification~\ref{spec:HGinvariant} obviously implies that the right-hand side in \eqref{eq:relationGHG} is constant for every $k\in\N$.
 Due to the invariant, this property directly translates to $\tilde{\Gb}_k \tilde{\Hb}_k^{-1} \tilde{\Gb}_k^\top$.
The proof for $\tilde{\Gb}_k \tilde{\Hb}_k^{-1} \tilde{\fb}_k + \tilde{\eb}_k$ being constant  for every $k\in \Kc$ under Specifications~\ref{spec:HGinvariant} and~\ref{spec:efkConstant} is analogous.
\end{proof}

Based on Lemma~\ref{lem:SpecsAndInvariants}, the cloud obtains necessary conditions for checking whether certain specifications apply. Before exploiting this feature in Section~\ref{sec:Attacks},
we briefly note that invariants similar to~\eqref{eq:invariants} also
arise if the permuted parameters $\Gbt_k^\prime$ and $\ebt_k^\prime$ (as specified in~\eqref{eq:permutedTransformations}) replace $\Gbt_k$ and $\ebt_k$ in~\eqref{eq:transformations}.
In fact, we then~find
\begin{subequations}
\label{eq:invariantsPermuted}
\begin{align}
\label{eq:relationPermutedGHG}
    \tilde{\Gb}_k^\prime \tilde{\Hb}_k^{-1} (\tilde{\Gb}_k^\prime)^\top  &= \Pb_k \Gb_k  \Hb_k^{-1} \Gb_k^\top \Pb_k^\top \qquad \text{and} \\
     \label{eq:relationPermutedGHfe}
    \tilde{\Gb}_k^\prime \tilde{\Hb}_k^{-1} \tilde{\fb}_k + \tilde{\eb}_k^\prime
   &= \Pb_k \left(\Gb_k \Hb_k^{-1} \fb_k + \eb_k \right).
\end{align}
\end{subequations}
Finally, also the set of active constraints (or, if permutations are involved, the set's cardinality) is invariant under~\eqref{eq:transformations}.

\subsection{Consistent and correct guesses}

Next, we focus on the systematic derivation of guesses (for $\Hbh_k$, $\Gbh_k$, $\fbh_k$, $\ebh_k$, $\Rbh_k$, and $\rbh_k$) that are consistent with the observed data according to~\eqref{eq:consistentGuess}.
It is easy to see that a trivial solution to this task is
\begin{subequations}
\label{eq:trivialGuess}
\begin{align}
 \Hbh_k&:=\Hbt_k, &\Gbh_k&:=\Gbt_k,&\Rbh_k&:=\Ib_l, \\
 \fbh_k&:=\fbt_k,&\ebh_k&:=\ebt_k, &\rbh_k&:=\zerob.
\end{align}
\end{subequations}
While trivial, the family of associated guesses~\eqref{eq:additionalGuesses} includes the original parameters as apparent from the (existing but typically unknown) transformation with $\Rbt_k:=\Rb_k^{-1}$ and $\rbt_k=-\Rbt_k\rb_k$, which indeed leads~to the correct guess
\begin{subequations}
\label{eq:correctGuess}
\begin{align}
 \Hbv_k&=\Hb_k, &\Gbv_k&=\Gb_k,&\Rbv_k&=\Rb_k, \\
 \fbv_k&=\fb_k,&\ebv_k&=\eb_k, &\rbv_k&=\rb_k.
\end{align}
\end{subequations}
Remarkably, any consistent guess can, in principle, be transformed into a correct guess via~\eqref{eq:additionalGuesses} as formalized in the following lemma.

\begin{lem}
Assume a consistent guess satisfying~\eqref{eq:consistentGuess} is known. Then, the transformation \eqref{eq:additionalGuesses} with $\Rbt_k:=\Rbh_k \Rb_k^{-1}$ and $\rbt_k:=\rbh_k-\Rbt_k \rb_k$ yields a correct guess satisfying~\eqref{eq:correctGuess}.
\end{lem}

\begin{proof}
 We initially note that $\Rbv_k=\Rb_k$ and $\rbv_k=\rb_k$ immediately results when substituting $\Rbt_k$ and $\rbt_k$ in~\eqref{eq:additionalRr}. Further, substituting $\Rbt_k$ in \eqref{eq:additionalHG} and using
 \eqref{eq:consistentGuessHG} and \eqref{eq:transformationsHG} proves $\Hbv_k=\Hb_k$ and $\Gbv_k=\Gb_k$. Analogously, we can prove $\fbv_k=\fb_k$ and $\ebv_k=\eb_k$ by exploiting~\eqref{eq:transformations_fe} and \eqref{eq:consistentGuess_fe}.
\end{proof}

While consistent with the data,
the trivial guess~\eqref{eq:trivialGuess} is not very helpful if Specification~\ref{spec:HGinvariant} applies
because, up until this point, consistency has only been defined with respect to isolated instances $k$
(typically leading to guesses that are not consistent across different instances, e.g., $\Hbh_0 \neq \Hbh_{k}$). In order to avoid this issue, we derive another consistent guess based on the invariant~\eqref{eq:relationGHG}, which promotes
cross-instance consistency according to Lemma~\ref{lem:SpecsAndInvariants}.
This novel guess builds on a singular value decomposition (SVD)
of the matrix on the left-hand side of~\eqref{eq:relationGHG}
 leading to
\begin{equation}
\label{eq:svd-invariant-tilde}
\Gbt_k \Hbt_k^{-1} \Gbt_k^\top  = \Ubt_k \Sigmabt_k \Vbt_k^\top,
\end{equation}
where $\Ubt_k :=\begin{pmatrix}
   \tilde{\ub}_{1,k} & \!\dots\! & \tilde{\ub}_{q,k}
\end{pmatrix} \in \R^{q \times q}$ and $\Vbt_k$ are orthogonal and where
\begin{equation*}
\tilde{\Sigmab}_k =\begin{pmatrix}
   \Dbt_k & \zerob \\
  \zerob  & \zerob
\end{pmatrix} \in \R^{q \times q} \,\, \text{with} \,\, \Dbt_k:=
\diag( \tilde{\sigma}_{1,k}, \dots,  \tilde{\sigma}_{l,k} ).
\end{equation*}
The singular values $\tilde{\sigma}_{1,k}\geq ... \geq \tilde{\sigma}_{l,k}>0$ are $l$ in number, since Assumption~\ref{assum:HposDefGRankl} immediately implies that $\Gb_k \Hb_k^{-1} \Gb_k^\top$ is of rank $l$. Due to~\eqref{eq:relationGHG}, the same applies to $\tilde{\Gb}_k \tilde{\Hb}_k^{-1} \tilde{\Gb}_k^\top$ yielding $l$ positive $\tilde{\sigma}_{i,k}$.
We then find the following theorem.

\begin{thm}
\label{thm:guessSVD}
The guess with the parameters
\begin{equation*}
  \Hbh_k:=\Dbt_k^{-1}\!,   \,\,\,\,\Gbh_k:=\begin{pmatrix}
    \tilde{\ub}_{1,k} & \dots &\tilde{\ub}_{l,k}
  \end{pmatrix},\,\, \text{and}\,\, \Rbh_k :=\Gbh_k^\top \Gbt_k
\end{equation*}
is consistent with~\eqref{eq:consistentGuessHG}.
\end{thm}

\begin{proof}
To prove the claim, we first note that
$\Gbh_k^\top \Ubt_k = \begin{pmatrix} \Ib_l & \zerob \end{pmatrix}$ by construction of $\Gbh_k$. Furthermore, since $\tilde{\Gb}_k \tilde{\Hb}_k^{-1} \tilde{\Gb}_k^\top$ is positive semi-definite with rank $l$, the first $l$ columns of $\Vbt_k$ are equivalent to those of $\Ubt_k$. Hence, $\Gbh_k^\top \Vbt_k = \begin{pmatrix}
\Ib_l & \ast
  \end{pmatrix}$.
Next, we multiply~\eqref{eq:svd-invariant-tilde} with  $\Gbh_k^\top $ from the left and with $\Gbh_k$ from the right to obtain
\[
\Rbh_k \Hbt_k^{-1}\Rbh_k^\top = \begin{pmatrix}
\Ib_l & \zerob
  \end{pmatrix} \Sigmabt_k \begin{pmatrix}
\Ib_l \\ \ast
  \end{pmatrix} = \Dbt_k.
\]
Since only square matrices are involved and $\Dbt_k$ is regular by construction, this relation implies invertibility of $\Rbh_k$ and
$\Dbt_k^{-1}=\Rbh_k^{-\top}\Hbt_k\Rbh_k^{-1}$.
With this at hand, we find
\[
\Rbh_k^\top \Hbh_k \Rbh_k = \Rbh_k^\top \Dbt_k^{-1} \Rbh_k = \Hbt_k,
\]
which proves the first equation in~\eqref{eq:consistentGuessHG}. It remains to prove $\Gbt_k=\Gbh_k \Rbh_k$. Hence, we define \mbox{$\Ybt_k:=\begin{pmatrix} \tilde{\ub}_{l+1,k} & \dots &   \tilde{\ub}_{q,k} \end{pmatrix}$}
and note that
$\Ybt_k^\top \Ubt_k = \begin{pmatrix}
    \zerob & \Ib_{q-l}
\end{pmatrix}$ and $
\Ybt_k^\top \Vbt_k =\begin{pmatrix} \zerob & \ast \end{pmatrix}$.
As a consequence, we find $\Ybt_k^\top \Ubt_k \Sigmabt_k \Vbt_k^\top \Ybt_k=\zerob$. Due to~\eqref{eq:svd-invariant-tilde}, this also implies $\Ybt_k^\top \Gbt_k \Hbt_k^{-1} \Gbt_k^\top \Ybt_k=\zerob$. Now, due to positive definiteness of $\Hbt_k^{-1}$, the latter relation holds if and only~if $\Ybt_k^\top \Gbt_k=\zerob$.
Furthermore, we have $\Gbh_k \Gbh_k^\top=\Ib_q - \Ybt_k \Ybt_k^\top$ due to $\Ubt_k \Ubt_k^\top = \Ib_q$. Hence, we obtain
\[
\Gbh_k \Rbh_k = \Gbh_k \Gbh_k^\top \Gbt_k  = (\Ib_q - \Ybt_k \Ybt_k^\top) \Gbt_k = \Gbt_k,
\]
which completes the proof.
\end{proof}

Theorem~\ref{thm:guessSVD} indicates that finding consistent guesses according to \eqref{eq:consistentGuess} can be decoupled. In fact, consistent $\Hbh_k$, $\Gbh_k$, and $\Rbh_k$ can be identified by only considering \eqref{eq:consistentGuessHG} and  $\fbh_k$, $\ebh_k$, and $\rbh_k$ only appear in \eqref{eq:consistentGuess_fe}.
In particular, once consistent $\Hbh_k$, $\Gbh_k$, and $\Rbh_k$ have been found, \eqref{eq:consistentGuess_fe} can be rewritten in terms of the linear equations
\begin{equation}
\label{eq:linearEqsFor_efrk}
    \begin{pmatrix}
        \Ib_q & \zerob & -\Gbh_k \\
        \zerob & \Rbh_k^\top & \Rbh_k^\top \Hbh_k
    \end{pmatrix} \begin{pmatrix}
        \ebh_k \\ \fbh_k \\ \rbh_k
    \end{pmatrix} = \begin{pmatrix}
        \ebt_k \\
        \fbt_k
    \end{pmatrix}
\end{equation}
with the unknowns  $\fbh_k$, $\ebh_k$, and $\rbh_k$. These equations are,
 e.g., solved by $\fbh_k:=\Rbh_k^{-\top} \fbt_k$, $\ebh_k:=\ebt_k$, and $\rbh_k:=\zerob$. However,  {having $q+l$ equations for $q+2l$ unknowns, it is immediately clear that} the system of equations is underdetermined and (infinitely many) more solutions exist. Moreover, the  {particular solution mentioned above} is typically inconsistent with the Specifications~\ref{spec:ekFixedUpperPart} and \ref{spec:efkConstant}. As a consequence, we investigate~\eqref{eq:linearEqsFor_efrk} in more detail in the next section.

\subsection{Exploiting linear dependencies of parameters}
\label{subsec:linearParameterDependencies}

Similar to Specification~\ref{spec:HGinvariant}, also Specifications~\ref{spec:ekFixedUpperPart} and \ref{spec:efkConstant} call for consistent guesses across multiple problem instances.
In the following, we will analyze the effect of these specifications on~\eqref{eq:linearEqsFor_efrk}, assuming throughout that Specification~\ref{spec:HGinvariant} applies as well. For ease of presentation, we begin by analyzing the effect of Specification~\ref{spec:efkConstant}.
In this context, we assume that the index set $\Kc$ contains $s$ elements and enumerate them by $k_1,\dots,k_s$.
Then, taking $\ebh_{k_1}=\dots=\ebh_{k_s}$ and $\fbh_{k_1}=\dots=\fbh_{k_s}$ into account, \eqref{eq:linearEqsFor_efrk} results in the following relation:
\begin{equation}
\label{eq:sysOfEquationsSpec3}
    \!\!\begin{pmatrix}
        \Ib_q  & \zerob & -\Gbh_0 &          & \zerob         \\
        \vdots & \vdots &   &  \!\!\ddots\!\! &                \\
        \Ib_q  & \zerob & \zerob  &   & -\Gbh_{0} \\
        \zerob & \Rbh_{k_1}^\top        & \!\!\Rbh_{k_1}^\top \Hbh_0 &                          & \zerob  \\
        \vdots & \vdots             &             & \!\!\ddots\!\!               &         \\
        \zerob & \Rbh_{k_s}^\top & \zerob             &         &  \Rbh_{k_s}^\top \Hbh_{0}
    \end{pmatrix} \begin{pmatrix}
        \ebh_{k_1} \\ \fbh_{k_1} \\ \rbh_{k_1} \\ \vdots \\ \rbh_{k_s}
    \end{pmatrix} = \begin{pmatrix}
        \ebt_{k_1} \\
        \vdots \\
        \ebt_{k_s} \\
        \fbt_{k_1} \\
        \vdots \\
        \fbt_{k_s}
    \end{pmatrix}\!.\!
\end{equation}
At this point, we note that based on $\Rbh_{k_1}$, $\Hbh_0,$ and $\Gbh_0$ (e.g., obtained via Theorem~\ref{thm:guessSVD}) one can construct the coefficient matrix in~\eqref{eq:sysOfEquationsSpec3} because the remaining keys $\Rbh_{k_2},\Rbh_{k_3},\dots,\Rbh_{k_s}$ follow from $\Gbh_0$ and $\Gbt_k$ with $k\in\Kc$.
Now, at first glance, this system of equations looks overdetermined with $(q+l)s$ equations and $q+l(s+1)$ unknowns. However, it is easy to show (but omitted here for brevity) that the rank of the matrix in~\eqref{eq:sysOfEquationsSpec3} is only $q+ls$. Hence, the system of equations~\eqref{eq:sysOfEquationsSpec3} is again underdetermined and $l$ (independent) equations are again missing. In contrast to~\eqref{eq:linearEqsFor_efrk}, however, where $l$ equations were missing for \emph{each} $k\in\N$, here $l$ equations are missing to identify all unknowns for \emph{all} instances $k\in\Kc$.

In addition, we can exploit that the simultaneous application of Specifications~\ref{spec:HGinvariant} and \ref{spec:efkConstant} also implies identical QPs~\eqref{eq:originalQP} for $k\in\Kc$ and consequently $\zb^\ast_{k_1}=\dots=\zb^\ast_{k_s}$.
We can make use of this relationship by noting that, once a guess $\hat{\zb}^\ast_{k_i}$ for any original optimizers $\zb^\ast_{k_i}$ is known, we can immediately derive $\rbh_{k_i}:=\hat{\zb}^\ast_{k_i}-\Rbh_{k_i} \yb_{k_i}^\ast$, $\ebh_{k_i}:=\ebt_{k_i}+\Gbh_0 \rbh_{k_i}$, and $\fbh_{k_i}:=\Rbh_k^{-\top} \fbt_{k_i}-\Hbh_0 \rbh_{k_i}$.

It remains to comment on the effects of Specification~\ref{spec:ekFixedUpperPart}. Although this specification addresses all $k\in \N$, we consider the same set $\Kc$ as above for convenience. In this context, \eqref{eq:linearEqsFor_efrk} leads to a system of linear equations similar to~\eqref{eq:sysOfEquationsSpec3} and also with the same right-hand side but with the unknown~vectors
\begin{equation}
\label{eq:unknowsSpec2}
\ebh^{\text{fix}}, \,  \ebh_{k_1}^{\mathrm{var}}, \,  \dots , \,  \ebh_{k_s}^{\mathrm{var}} , \,   \fbh_{k_1}, \,  \dots , \, \fbh_{k_s} , \, \rbh_{k_1}, \, \dots , \,  \rbh_{k_s}.
\end{equation}
Hence, we (initially) have $q_{\text{fix}}+(q_{\text{var}}+2l)s$ unknowns here but again $(q+l)s$ equations. Now, assuming
\begin{equation}
\label{eq:condition_qfix}
    q_{\text{fix}}>l \qquad \text{and} \qquad s \geq \frac{q_{\text{fix}}}{q_{\text{fix}}-l},
\end{equation}
the number of equations exceeds (or matches) the number of unknowns. However, even if~\eqref{eq:condition_qfix} holds, one can show that the rank deficiency of the coefficient matrix characterizing the system of linear equations is again~$l$. Still, once we have a guess for either $\ebh^{\text{fix}}$ or any of the $\fbh_{k_i}$ or $\rbh_{k_i}$, we can compute all unknowns~\eqref{eq:unknowsSpec2}.
We make use of this feature in Section~\ref{sec:Attacks}.

\subsection{Partially resolving permutations}
\label{subsec:permutations-analysis}
As apparent from~\eqref{eq:invariantsPermuted}, the invariants~\eqref{eq:invariants} exploited above lose some of their informative value if permutations are involved.
Nevertheless, if Specification~\ref{spec:HGinvariant} applies, the entries of $\Gbt_k^\prime \Hbt_k^{-1} (\Gbt_k^\prime)^\top$ and $ \Gbt_0^\prime \Hbt_0^{-1} (\Gbt_0^\prime)^\top$ differ only in their position.
Thus, whenever these matrices contain at least~$q$ distinct entries, which is often the case, it is straightforward to identify relative permutations $\Delta P_{k,0}:=P_{k} P_{0}^\top$
such that
\begin{equation*}
\tilde{\Gb}_{k}^\prime \tilde{\Hb}_{k}^{-1} (\tilde{\Gb}_{k}^\prime)^\top =
\Delta P_{k,0}
\tilde{\Gb}_{0}^\prime \tilde{\Hb}_0^{-1} (\tilde{\Gb}_{0}^\prime)^\top \Delta P_{k,0}^\top.
\end{equation*}
Moreover, if also Specification~\ref{spec:efkConstant} applies, the identification can even be simplified based on relations like
\begin{equation*}
\tilde{\Gb}_{k}^\prime \tilde{\Hb}_{k}^{-1} \tilde{\fb}_{k} + \tilde{\eb}_{k}^\prime =
\Delta P_{k,0} \big(\tilde{\Gb}_0^\prime \tilde{\Hb}_0^{-1} \tilde{\fb}_0 + \tilde{\eb}_0^\prime\big).
\end{equation*}
Now, suppose
that $P_0$ (or any other ``absolute'' permutation) is known. Then
one can resolve all other permutations by evaluating $P_k=\Delta P_{k,0} P_0$.
This eventually allows computing $\Gbt_k=P_k^\top \Gbt_k^\prime$ and $\ebt_k=P_k^\top \ebt_k^\prime$ such that we deal with the unpermuted problem again.

\section{Attack scenarios}
\label{sec:Attacks}

In this section, we consider an exemplary application of the transformed QP~\eqref{eq:transformedQP} in the context of privacy-preserving MPC and illustrate how our theoretical findings about the transformation itself and the transformed
parameters~\eqref{eq:transformations} can be used to launch attacks. The control scheme will be applied to a mobile robot,
for which we briefly clarify the corresponding system model, the MPC setup, and the cloud-based solution of the resulting QP next.

\subsection{Specification and cloud-based realization of MPC}

\begin{figure}[t]
    \centering
    \includegraphics[width=\columnwidth]{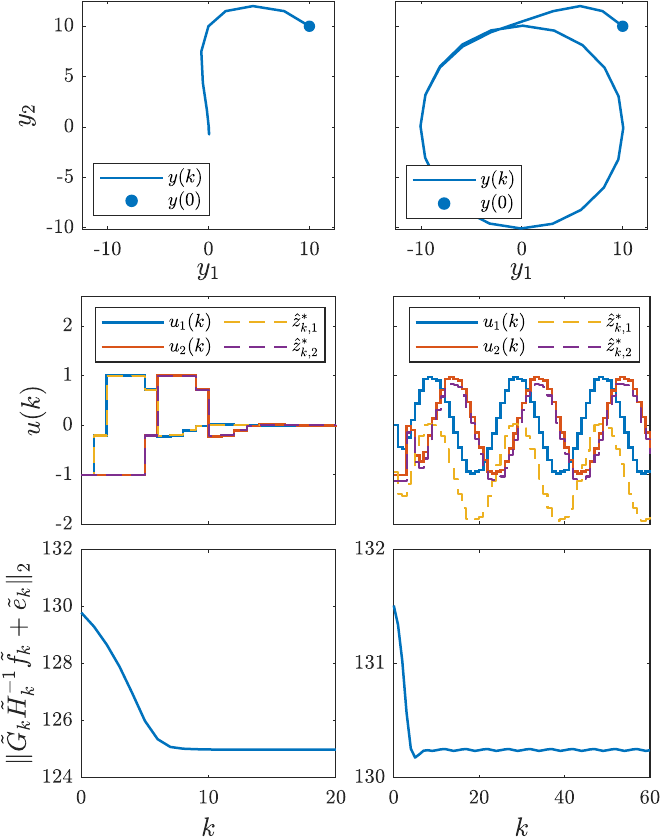}%
    \caption{Output trajectories (top), input sequences (middle), and norms of~\eqref{eq:relationGHfe} (bottom) for the first (left, setpoint) and second (right, tracking) experiment, respectively.}%
    \label{fig:setpoint-and-tracking}
    \vspace*{-5mm}
\end{figure}

We assume linear discrete-time system dynamics modeled by $\xb(k+1)=\Ab \xb(k) + \Bb \ub(k)$ with
\begin{equation*}
\Ab:=\Ib_2 \otimes \begin{pmatrix}
    1 & 1 \\
    0 & 1
\end{pmatrix} \quad \text{and} \quad \Bb:=\Ib_2 \otimes \begin{pmatrix}
    0.5 \\
    1
\end{pmatrix},
\end{equation*}
where ``$\otimes$'' denotes the Kronecker product.
The system is subject to the state and input constraints
$\underline{\xb} \leq \xb(k) \leq \overline{\xb}$ and $\underline{\ub} \leq \ub(k) \leq \overline{\ub}$, respectively, with
\begin{equation}
\label{eq:stateInputConstraintsExample}
    \!\!\overline{\xb}=-\underline{\xb}:=\begin{pmatrix}
    20 & \!5 & \!20 & \!5
\end{pmatrix}^\top  \text{and} \,\,\, \overline{\ub}=-\underline{\ub}:=\begin{pmatrix}
    1 & \!1
\end{pmatrix}^\top\!\!\!.\!
\end{equation}
In each time-step $k$, we
minimize the cost function
\begin{equation*}
\sum_{\kappa=k}^{k+N-1} \|\yb(\kappa)-\yb_{\text{ref}}(\kappa) \|_{\Qb}^2 +\|\ub(\kappa) - \ub(\kappa-1)\|_{\Rb}^2
\end{equation*}
subject to the dynamics and constraints from above for $N=5$ prediction steps,
where the outputs \mbox{$\yb(k):=(\xb_1(k) \;\; \xb_3(k))^\top$} reflect the robot's position.
The weighting matrices are chosen as $\Qb=\Ib_2$ and $\Rb=0.1 \Ib_2$. As for the references, we will first consider ${\yb_{\text{ref}}(k)=\zerob}$ reflecting a setpoint and then
a sinusoidal reference representing a circle with radius $10$ traversed in counterclockwise direction with a period
of $20$ time steps.
For both numerical experiments,
the robot's initial state is
\mbox{$\xb(0)=\left(10 \; -2 \;\; 10 \;\; 2 \right)^\top$} and $\ub(-1)=\zerob$.
Now, reformulating the control task in terms of the QP~\eqref{eq:originalQP} is straightforward, and we omit details for brevity. We just note that (after condensation) the decision variables $\zb$ reflect the predicted input sequences in terms of ${\ub(k),\dots,\ub(k+N-1)}$.
Furthermore, for every $k \in \N$, the QP parameters~\eqref{eq:parametersQP} can be stated~as
\begin{subequations}%
\label{eq:HGefExample}%
\vspace*{-5ex}%
\begin{align}
  \label{eq:HGeExample}
   \!\!\!H_k\!&=H_0, \,\,\, G_k=G_0:=\begin{pmatrix}
    \blind{-}\Ib_{l} \\
    -\Ib_{l} \\
    \ast
\end{pmatrix}\!, \,\, \eb_k:= \begin{pmatrix}
    \blind{-}\overline{\zb} \\
    -\underline{\zb} \\
    \varepsb(\xb(k))
    \end{pmatrix}\!,\!\!\\
    \!\!\!\fb_k&:=\varphib(\xb(k),\ub(k-1),\yb_{\text{ref}}(k),\dots, \yb_{\text{ref}}(k+N-1))\!\!
\end{align}%
\end{subequations}
with $\varepsb:\R^n \rightarrow \R^{2nN}$ and $\varphib:=\R^{n+m+Np} \rightarrow \R^l$, where ${m=2}$, ${n=4}$, and ${p=2}$ reflect the input, state, and output dimension of the system at hand and where the number of decision variables is $l=mN=10$.
The constraint vectors $\underline{\zb}$ and $\overline{\zb}$ result from stacking the input constraints in~\eqref{eq:stateInputConstraintsExample}.
Finally, in each time step, we apply the
optimal input $\ub^\ast(k)$ (the first $m$ entries of $\zb_k^\ast$) to the system and repeat the procedure at the next sampling instant $k+1$.
For an overview, the resulting system trajectories and inputs for both references are illustrated in the upper and middle charts of Figure~\ref{fig:setpoint-and-tracking}, respectively.

Finally, a cloud-based solution of the QPs via~\eqref{eq:transformedQP} is realized as follows.
First, we randomly and independently choose $\Rb_k$ and $\rb_k$
for each $k$. Since suitable distributions are typically not specified in the corresponding literature, we sample floating point numbers uniformly from the interval $[-10,10]$ for simplicity here.
Then, the cloud receives the transformed parameters $\Hbt_k$, $\Gbt_k$, $\ebt_k$, and $\fbt_k$ (computed according to~\eqref{eq:transformations}), solves \eqref{eq:transformedQP}, and returns $\yb_k^\ast$ to the client, where $\zb_k^\ast$ is recovered via~\eqref{eq:relationOfOptimizers}.

\subsection{Preparing and launching attacks}
\label{subsec:attacks-experiments}

Summarizing Section~\ref{sec:problemAnalysis}, we rely mostly on the relations between the transformed and original parameters stemming from Specifications~\ref{spec:HGinvariant} to \ref{spec:efkConstant} here.
Via Lemma~\ref{lem:SpecsAndInvariants}, we can see that constant $\tilde{\Gb}_k \tilde{\Hb}_k^{-1} \tilde{\Gb}_k^\top$ (for all $k\in\N$) and $\tilde{\Gb}_k \tilde{\Hb}_k^{-1} \tilde{\fb}_k + \tilde{\eb}_k$ (for all $k$ in some set $\Kc\subseteq \N$) are strong indicators for the validity of
Specification~\ref{spec:HGinvariant} and~\ref{spec:efkConstant}, respectively. Here (due to~\eqref{eq:HGeExample}), the cloud will observe constant $\tilde{\Gb}_k \tilde{\Hb}_k^{-1} \tilde{\Gb}_k^\top$ for all $k$ in both experiments and will thus (correctly) assume that Specification~\ref{spec:HGinvariant} applies.
Regarding the second invariant, we depict $\|\tilde{\Gb}_k \tilde{\Hb}_k^{-1} \tilde{\fb}_k + \tilde{\eb}_k\|_2$ in the bottom charts of Figure~\ref{fig:setpoint-and-tracking}. Remarkably,
because $\fb_k$ and $\eb_k$ depend on the current state,
some properties of the system's behavior are leaked inevitably. For instance, the convergence time to the reference and, in case of the tracking problem, the fact that the motion is periodic are revealed.
Moreover, $\|\tilde{\Gb}_k \tilde{\Hb}_k^{-1} \tilde{\fb}_k + \tilde{\eb}_k\|_2$ is (almost) constant for all $k\in\Kc_1:=\{8,9,\dots,20\}$ in the first experiment and for various triplets such as $\{15, 20, 25\}$ or $\Kc_2:=\{10, 30, 50\}$ in the second one.
A closer investigation of the involved vectors reveals that in the first experiment also $\tilde{\Gb}_k \tilde{\Hb}_k^{-1} \tilde{\fb}_k + \tilde{\eb}_k$ is constant for all $k \in \Kc_1$, and the cloud will thus (correctly) assume that Specification~\ref{spec:efkConstant} applies at these time steps. Regarding the second experiment, the situation is slightly more complicated. In fact, while the norm is (almost) constant for both of the above triplets, the vectors $\tilde{\Gb}_k \tilde{\Hb}_k^{-1} \tilde{\fb}_k + \tilde{\eb}_k$ vary for the first triplet but are (almost) constant for $k\in\Kc_2$. Based on these observations, the cloud not only gets an indication for the validity of Specification~\ref{spec:efkConstant} during some time steps for the second experiment, but it can even infer the (correct) reference period of $20$ time steps.

Now, in order to launch an actual attack aiming for a reconstruction of the original QP parameters, the cloud could make use of Theorem~\ref{thm:guessSVD} to derive consistent guesses for $\Hb_0$, $\Gb_0$, and all  $\Rb_{k}$. However, due to the special but common structure of $\Gb_0$ from~\eqref{eq:HGeExample}, the cloud can obtain a more powerful guess more easily.
In fact, it will receive the transformed matrices $\Gbt_k^\top = \begin{pmatrix} \Rb_k^\top & -\Rb_k^\top & \ast \end{pmatrix}$ and observe that the second square block is always a negation of the first one. Having already a strong indication for a constant $\Gb_k=\Gb_0$ and likely knowing about the common structure in MPC, it will simply pick the first square block of $\Gbt_k$ as the (correct) guess $\Rbh_k$. Immediately, this leads to the (likewise correct) guesses $\Gbh_0=\Gbt_k\Rbh_k^{-1}$ and $\Hbh_0=\Rbh_k^{-\top} \Hbt_k \Rbh_k^{-1}$.

In order to completely break the cipher, it remains to reconstruct $\eb_k$, $\fb_k$, and $\rb_k$. Hence, the cloud builds up the system of equations~\eqref{eq:sysOfEquationsSpec3} for the steps in
$\Kc_i$ with $i\in\{1,2\}$.
Now, as pointed out in Section~\ref{subsec:linearParameterDependencies}, it is necessary to add $l$ additional equations to avoid that the system is underdetermined.
As proposed, the cloud will add a guess for the constant $\zb_k^\ast$ during the instances
in $\Kc_i$, respectively. For the first experiment, the lower left chart in Figure~\ref{fig:setpoint-and-tracking} suggests that the system converges to an equilibrium.
Hence, $\zbh_k^\ast=0$ is a reasonable guess here. In the absence of more reasonable alternatives, the cloud will use the same guess for the second example, although it is most likely erroneous there.
After adding $\zbh_k^\ast=0$, the cloud solves~\eqref{eq:sysOfEquationsSpec3} and obtains $\ebh_k$, $\fbh_k$, and $\rbh_k$ for all $k\in \Kc_i$.

Finally, to also address the steps not contained in the sets $\Kc_i$, the cloud (correctly) assumes that Specification~\ref{spec:ekFixedUpperPart} holds, which is reasonable based on the assumed structure of $\Gbh_0$. In fact, the blocks $\Ib_l$ and $-\Ib_l$ typically refer to constant box constraints as in~\eqref{eq:zBoxConstraints}. This corresponds to $q_{\text{fix}}= 2l$ implying that the first condition in~\eqref{eq:condition_qfix} is satisfied and that the second holds whenever $s\geq 2$. Hence, the cloud straightforwardly includes all remaining time steps $k$
in the system of equations for the unknowns~\eqref{eq:unknowsSpec2}
and adds an already solved one from $\Kc_i$ to ensure that the system is determined.
After solving for these unknowns, all $\zbh_k^\ast$ can be computed via~\eqref{eq:relationOfOptimizers}.
The first two components of $\zbh_k^\ast$ are illustrated in the middle charts of Figure \ref{fig:setpoint-and-tracking}. As apparent, the inputs applied to the system are accurately reconstructed for the first experiment. In the second experiment, despite the erroneous guess $\zbh_k^\ast=0$ for $k\in \Kc_2$, only a constant offset is present in the reconstructed input sequences (the shape of the signals is accurately recovered).
Note that the magnitude of the error in each component varies with the choice of $\zbh_k^\ast$. Since $\zb^\ast_{k,2}\approx 0 = \zbh_{k,2}^\ast$ for all $k\in \Kc_2$, the second component of the inputs is almost correctly reconstructed here.

\subsection{Scenarios with permutations}
\label{subsec:attacks-permutations}

We briefly address the case, where the transformations are accompanied by permutations according to~\eqref{eq:permutedTransformations}. While this simple modification significantly complicates a reconstruction, we pointed out in Section~\ref{subsec:permutations-analysis} that relative permutations such as $\Delta \Pb_{k,0}$ can often be identified with moderate effort.
This is indeed the case in our experiments,
even though the special structure of $\Gb_0$ (together with the choices for $\Qb$ and $\Rb$) leads to ambiguous entries in $\Gbt_k^\prime \Hbt_k^{-1} (\Gbt_k^\prime)^\top$.
Hence, picking up the assumption from Section~\ref{subsec:permutations-analysis} that $\Pb_0$ is known, we can resolve all permutations for the experiments at hand and subsequently handle the reconstruction of the QP parameters as before.
Regarding the identification of an unknown $\Pb_0$, we note that the special structure of $\Gb_0$  excludes many realizations of $\Pb_0$. However, without additional knowledge, the problem of uniquely identifying $\Pb_0$ is intractable.

\subsection{Remarks on the attacks and additional knowledge}

We conclude this section with some remarks on the proposed attacks. As apparent from Figure~\ref{fig:setpoint-and-tracking}, the cloud is, in principle, able to infer crucial characteristics of the control system and to recover the majority of the sensitive data. However, the cloud is currently not able to verify its guesses without additional knowledge.
In fact, this corresponds to a ciphertext-only setup which is assumed in the literature.
Still, having suitable additional knowledge (e.g., exactly knowing that $\Gb_k$ or $\eb_k$ have the structure in~\eqref{eq:HGeExample}) seems reasonable for many real-world applications.
Further, it is in line with Kerckhoffs' principle~\cite[Sect.~1.2]{katz2007}, which (translated to this setup) states that the security of a cipher should not depend on an attackers' knowledge about the control system.
Especially, knowledge about the plant, which has been excluded here, is well-suited to validate the guesses from Section~\ref{subsec:attacks-experiments} resulting in a more targeted attack.

Furthermore, the most important observation resulting from our analysis is that only little information, which might even be public or easy to obtain, can enable to completely break RT ciphers in the context of QPs. Remarkably, this major issue is not present in other cryptosystems (such as homomorphic encryption~\cite{chillotti2020tfhe,Cheon2017_CKKS}), which could be used to ensure a privacy-preserving evaluation of QPs at the price of a higher computational load.

\section{Conclusions and Outlook}
\label{sec:Conclusions}

This paper deals with the security of random affine transformations in the context of model predictive control, where it is used for the private evaluation of quadratic programs. We show that the arising ciphertexts still contain information that can be exploited for an attack and, most importantly, that little additional information can suffice to break all ciphertexts even though keys are not reused in our setup.

In future work, additional equality constraints (as in~\cite{Zhou2015QPoutsourcing}) and a more detailed treatment of permutations are of interest. In this context, system knowledge, closed-loop effects, and the implementation over floating point numbers (see~\cite{IFAC2023_RandomTrafo}), entail valuable angles of attacks that we neglected here.

\section*{Acknowledgment}
Financial support by the German Research Foundation and the Daimler and Benz Foundation under the grants SCHU 2940/4-1 and 32-08/19 is gratefully acknowledged.

\end{document}